# DETERMINING THE NETWORK THROUGHPUT AND FLOW RATE USING GSR AND AAL2R

Adyasha Behera[1] and Amrutanshu Panigrahi[2]

[1]Department of Information Technology, College of Engineering and Technology, Bhubaneswar, India
[2] Department of Information Technology, College of Engineering and Technology, Bhubaneswar, India

## ABSTRACT

*In multi-radio wireless mesh networks, one node is eligible to transmit packets over multiple channels to different destination nodes simultaneously. This feature of multi-radio wireless mesh network makes high throughput for the network and increase the chance for multi path routing. This is because the multiple channel availability for transmission decreases the probability of the most elegant problem called as interference problem which is either of interflow and intraflow type. For avoiding the problem like interference and maintaining the constant network performance or increasing the performance the WMN need to consider the packet aggregation and packet forwarding. Packet aggregation is process of collecting several packets ready for transmission and sending them to the intended recipient through the channel, while the packet forwarding holds the hop-by-hop routing. But choosing the correct path among different available multiple paths is most the important factor in the both case for a routing algorithm. Hence the most challenging factor is to determine a forwarding strategy which will provide the schedule for each node for transmission within the channel. In this research work we have tried to implement two forwarding strategies for the multi path multi radio WMN as the approximate solution for the above said problem. We have implemented Global State Routing (GSR) which will consider the packet forwarding concept and Aggregation Aware Layer 2 Routing (AAL2R) which considers the both concept i.e. both packet forwarding and packet aggregation. After the successful implementation the network performance has been measured by means of simulation study.*

## 1. INTRODUCTION

A wireless mesh network (WMN) is a communication network consisting of radio nodes organized in a mesh topology [1]. Wireless mesh networks (WMNs) are becoming new and promising technique for internet connection due to its elegant feature like low cost for deployment, ease of use, long time network life, easy maintenance and robustness. Mesh router, mesh client and the gateways are three measure components of WMN. Routers are static in nature and provide the network backbone, the mesh clients are used to access the network through the mesh routers and also they directly mesh with each other and the gateways are used by the router for inter network communication. Some aspects like self- organising and self configuring of WMN make it different form the traditional wireless network [2]. In contrast WMN is a self sustaining network due to the above said features.

The routing protocol in a wireless mesh network is in charge of routing packets in such a way not to exceed the given available bandwidth on each link. Routing is the process of determining the path from a source to a destination. The main objective of routing scheme is to maintain the QOS of the end user with the optimizing way of utilizing the network resources. But to achieve these above said goals is so easy due to some tradeoffs like black hole, routing loop, maximum bandwidth problem and maximum flow problem. In hop by hop communication if one of the intermediate hops goes down permanents and it is not able to forward any message to the next hop, resulting the black hole in the network. Routing loop itself suggests the problem by its name itself i.e., the message is not being forwarded to the destination as its being looped inside the network. According to the maximum bandwidth problem the routing scheme has to carry maximum number of traffic from source to destination while keeping the bandwidth satisfied. The maximum flow rate problem is to find the maximum flow from a source to destination with a given constant bandwidth. To overcome these tradeoffs the routing scheme has to consider some metric such as Expected Transmission Count (ETX), Expected Transmission Time (ETT), energy consumption and the path availability and reliability. Unfortunately, traditional destination-based routing protocols do not take into account the link bandwidth availability resulting from a given channel assignment and route packets along the shortest paths computed by using certain link metrics. Finding a set of link costs such that a given set of traffic demands are routed so that the link available bandwidths are not exceeded is a difficult problem. Also, such a solution would be tightly coupled to a particular set of traffic demands and the network performance may decrease as the traffic demands vary.

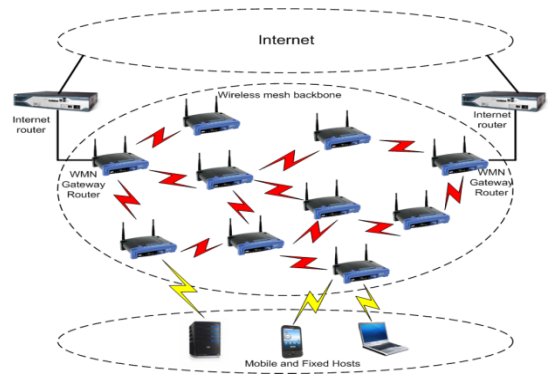

Figure 1. Infrastructure backbone of WMN

To overcome the shortcomings of traditional destination-based routing protocols, a new Layer-2.5 forwarding paradigm was proposed. In L2.5 [3], forwarding decisions are not taken by looking up the routing table, but are based on two objectives: i) balance the traffic among the outgoing links in proportion to their available bandwidth; ii) guarantee that all the packets reach the destination in a predetermined maximum number of hops.

The objective of our work is to detail study of the packet aggregation and packet forwarding strategy algorithms. Global state routing [4] and Aggregation Aware Layer 2.5 Routing (AAL2R) [25] algorithm are two packet forwarding strategies. These are tried to implement by using NS3 simulation. Environmental are setup for AAL2R and global state routing according to requirement specifications. Performance evaluation of WMN using global routing and AAL2R is our main focus. To achieve that we have calculated some of the network influencing parameter like network throughput, packet delivery ratio, packet loss ratio etc.

The paper is structured as follows. In chapter 2 a detailed literature study has been done on different available link matrices, packet forwarding and routing schemes. In 3 we have elaborated the packet forwarding and packet aggregation and described the AAL2R and GSR algorithm. In 4 a simulation study has been done in order to measure the network performance. Finally in 5 we have concluded for this research paper.

## 2. LITERATURE STUDY

In [5] the authors has introduced Expected Transmission Count (ETX) to estimate the number of transmission. The authors in [6] developed two metrics, the expected transmission time (ETT) and the weighted cumulative ETT (WCETT). MIC (metric of interference and channel switching) takes the interflow interference into account in addition to the intra-flow interference. A link metric based on the estimated available bandwidth is intended to be used with a single path destination-based routing protocol. In [7 and [8] authors had proposed two routing protocol named AODV and OLSR respective in order to consider the above link metric into consideration for ad-hoc network. These above said two routing protocols are single path routing protocols. The routing protocol specified in the IEEE 802.11s is basically a modified version of AODV that make the utilisation of the airtime link metric to associate each link with an estimated amount of time for successfully completion of transmission. These single path routing protocols follows some difficulties like limited capability and load balancing in the deployment phase. AODV-BR [9], AOMDV [10] are the extension of the single path routing protocols by considering multiple path between source and destination pair. Both finding considers the interference in some aspect, but don't consider the bandwidth constraint resulting from the channel assignment.

R. Draves *et.al. in* [22] has developed adaptive load-aware routing scheme and according to which the network is divided into multiple cluster. One cluster head is present for each cluster which is responsible for controlling the communication of the nodes belonging to its own cluster. A number of approaches exploit the broadcast nature of the wireless medium. ExOR [12] and GATOR [13] are two opportunistic approaches. In ExOR the node broadcasts a packet and the intermediate node receiving the packet will decide the next hop for further forwarding. This process will be iterated until the packet arrives at the proposed; however the protocol used to reach such agreement introduces some overhead. GATOR exploits the knowledge of geographic coordinate of the intermediate nodes while selecting the node for receiving the packet for further packet forwarding. But the drawback of the opportunistic approach is that it only considers the neighbouring nodes that are listening on the channel of the sender are eligible for receiving the packets. The authors in [13] has developed the ROMER which will result a mesh with minimum path cost and each packet is allowed to travel by any path from that mesh only. The any path routing paradigm [15] generalizes the opportunistic approach according to which every node is pre computed with the set of next-hops with different priority level. A packet is allowed to be forwarded to the highest priority next-hop and next hops are determined in decreasing order of priority. In [16] the authors have proposed one forwarding strategy to find the least cost any path. But Arun Raj *et. al.* [17] has extended the finding of [16] by considering more facts. First any path routing requires a modified MAC to determine which next hop has to forward the packet. Secondly, the load balancing need to be considered in case of failure of one node due to excessive load. Stefano Avallone et al. [5] had proposed a layer 2.5 algorithm which solves the link flow rates during channel assignment. In this L2.5, forwarding decisions are not taken by looking up the routing table, but are based on two objectives i) balancing the traffic on each outgoing links; ii) successful transmission of packets from source to destination must be constrained to a predetermined maximum number of hops. Arun Raj et al. [20] have described RSAPS (Round robin based Secure Aware Packet Scheduling) which will provide the forwarding strategy and also considers dynamically increasing or decreasing the security levels of packets based on the incoming load. RSAPS gives priority to schedulability rather than security. Stefano Avallone and Giovanni Di Stasi et al [21] has introduced the MPLS splitting policy requires to identify a suitable set of paths for each ingress-egress pair and to compute the set of split ratios.

## 3. IMPLANTATION OF ROUTING ALGORITHMS
**Packet Aggregation and Packet Forwarding**

Packet aggregation increases the network capacity by transmitting several packets at a single time interval. This technique sound to be elegantly effective when the overhead for a single transmission is high in the network. The benefits of packet aggregation have been shown in [22], [23], [24] and [25]. The recent IEEE 802.11n standard has also taken aggregation into consideration to improve performance. Deploying packet aggregation in multi-channel multi-path environments may result in suboptimal performance. This is because typically multi-path routing algorithms are also responsible to send data packets to different nodes simultaneous, but the aggregation concepts comes into account when multiple packet need to be forwarded to one intended recipient. Collection of different packets that are aimed to be transferred to one node is performed by the aggregation module. The aggregation module stores the packets into different queue. Each queue is for one next hop node that can be reached through the network interface. When a packet is received by the aggregation module, it is immediately time stamped and put into the appropriate queue of the next-hop the packet is destined to. The time stamp is used to determine long the packet has been queued already.

**Packet Forwarding**

The forwarding mechanism is nothing else the hop by hop routing. In packet forwarding one node just forwards the data packet to the next hop by looking into its routing table. When a packet arrives at one node it determines to keep the packet if that data packet is destined to it only otherwise it simply forwards the packet to the next of by getting the information from its routing table. By following this method iteratively packet is destined to the intended recipient. Due to the independent flow of packet inside the network the path for data flow need not be pre computed and the network is said to be connectionless [11], [25]. In order to properly route a packet, a router must be able to determine the next hop for the packet. For this purpose the router builds the routing table by considering the information returned by the routing protocol. The routing table should be dynamically updated in order to successfully handing the traffic. The table at each router is responsible for identifing the next hop for all known IP destination addresses. Routers generally store IP prefixes rather than complete IP addresses in their forwarding tables [4].

### 3.1. AAL2R

In Aggregation Aware Layer 2 Routing algorithm follow the packet aggregation as well as the forwarding strategy. All potential queues related to next-hop. The aggregation of the packet should be done by selecting one queue while maintaining the constant flow rate for the network. Finding the perfect transmission unit the packets follow the condition that the transmission unit must be non empty set and secondly the spare space (SP) should greater than the packet size. (SP) is defined as the interval between the transmissions of two packets. Hence SP is determined by subtracting header size and the number of packets in one queue from the maximum transmission unit.

$$SP = MTU - \sum_{p \epsilon Q_i} P_{size} - \text{Header Size} \quad (1)$$

Other condition which can arise during the transmission is the presence of empty set. In this case the aggregation set must wait for the upcoming packets for scheduling the transmission. The aggregation set is said to be empty in 2 cases. In 1st case when the queues are empty or the queues which are not empty and do not have enough spare space available for aggregation then the queue remains unchanged and can potentially aggregate a packet when a packet will arrive at the queue. In 2nd case if all queues hold some packets with a spare space smaller than the packet size then all the queues belonging to the candidate next-hops are considered eligible for sending the current packet. When the eligible set of next-hops has been chosen the queue is ready to transmit. If more than one queue are ready for transmission at the same time then the queue with the oldest packet is served first to avoid starvation. The oldest packet is determined by checking the time stamp attached with each packet as the time stamp shows the time that the packets has spent in queue for a chance of transmission. According to this the packet which holds the highest time stamp will have the highest priority for the transmission as the time stamp. By choosing one queue for transmission all other queues inside the aggregation set will be remain unchanged. Hence all queues are extracted from the aggregation set in the decreasing order of priority while the priority of the queue is determined by the average time stamp for each packet inside it.

### 3.2. Global State Routing

The Global State Routing Protocol (GSR) is aimed to determine the next hop of one node that wants to transmit the data into the wireless mesh network. The GSR also avoids the disadvantages of flooding of data into the network by taking link state table which will dynamically respond to the topology change inside the network and also periodically exchange the information with the neighbour node in order to maintain connectivity. The network is modelled as an undirected graph $G = (V, E)$, where V is a set of |V| j nodes and E is a set of |E| undirected links connecting nodes in V. In the network the each node has a unique identifier and represents a host with the transmission range and an undirected link connecting two nodes i and j is formed when the distance between i and j become less than or equal to transmission range. Link is removed from the graph when both nodes move away from each other. For each node i, one neighbour list, topology table, next hop table and distance table are maintained. Neighbour list contains the list of adjacent node currently available to the corresponding node. The topology table contains the link state information and the time stamp as well for the node. The next hop table contains the adjacent nodes called as next hop to which the node can transmit the data for the destination node. The distance table holds the shortest path between the source and destination node and it will be getting update after every successful transmission of data to a node from a node. Since min-hop and shortest path are only the two objective for this routing algorithm. A weight function is used to compute the distance of a link for each node, this weight function simply returns 1 if two nodes have direct connection, otherwise, it returns 0. For every node the weight function is applied to find out the shortest next hop from available multiple hops before the transmission of the data packet into the network.

## 4. SIMULATIONS AND RESULTS

In the following section our goal is to evaluate the packet forwarding strategies. And also calculate the network parameters like throughput and packet delivery ratio to determine the network performance. For this purpose we have used Network Simulator-3, which is an event driven simulator used for calculating the network performance. We have set different number of nodes and established the communication link between them. Then the UDP and TCP type traffic has been used over the link while communication. The above environment has been simulated in different time periods such as 10sec-60sec.

**Environment configuration**

To determine the performance of the algorithms by calculating some network parameters like packet delivery ratio, packet loss ratio, overall throughput. Simulation is done under the simulator version NS 3.20. In Figure 2 shows the arrangement of 10 nodes in ns3 environment Packet length taken 512 KB.

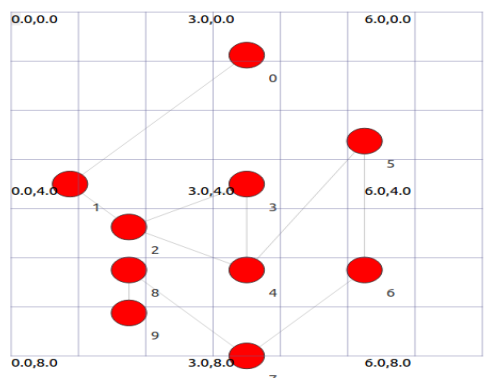

Figure 2. Node Setup in NS 3

**Packet Delivery Ratio**

Packet delivery ratio is defined as the ratio of data packets received by the destinations to those generated by the sources. In figure 3 shows the ratio of the number of delivered data packet to the destination. This illustrates the level of delivered data to the destination. Mathematically represented as

$$\text{Packet delivery ratio} = \frac{\sum \text{number of packet recived}}{\sum \text{number of packet send}}$$

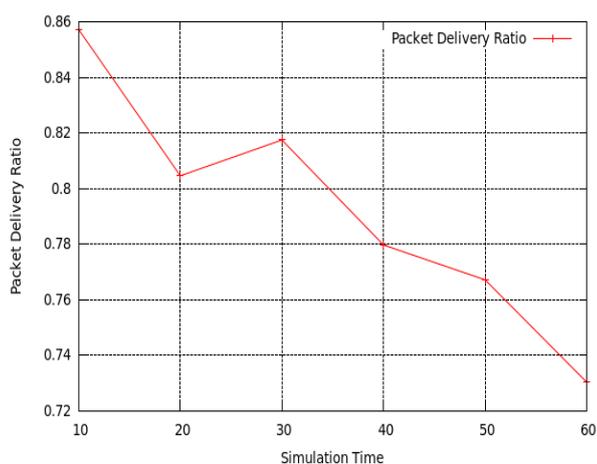

Figurer 3. Packet Delivery Ratio of GSR

**Packet Loss Ratio**

The packet loss ratio defined as deference of total number of packet send and Number of packet received. The figure 4 shows total number of packets dropped during the simulation a sample path for the delay and loss of the probe packets. The fluctuation in the graph indicates that there is a correlation between the loss and delay even in the actual network. Packet lost Ratio =∑ Number of packet send – ∑Number of packet received.

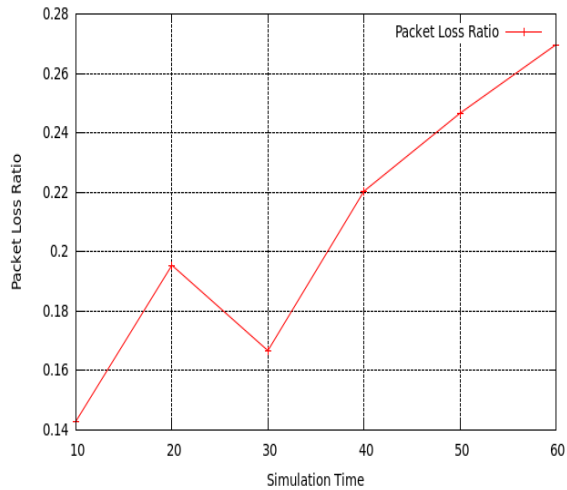

Figure 4. Packet Loss Ratio of GSR

**Average Throughput**

When used in the context of communication networks, such as Ethernet or packet radio, throughput or network throughput is the rate of successful message delivery over a communication channel. Figure 5 shows the data these messages belong to may be delivered over a physical or logical link or it can pass through a certain network node. Throughput is usually measured in bits per second (bit/s or bps), and sometimes in data packets per second or data packets per time slot. It is defined as the total number of packets delivered over the total simulation time.

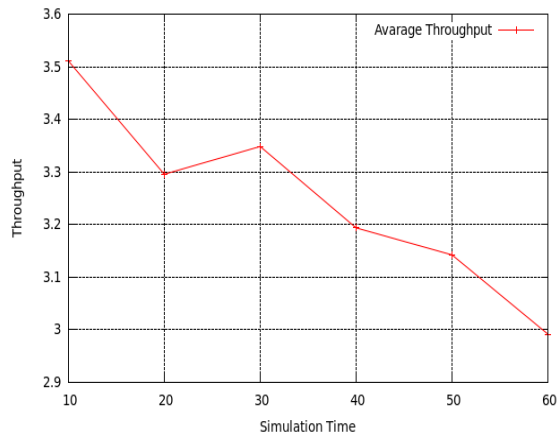

Figure 5. GSR Throughput in mbps

We evaluated the behaviour of the considered forwarding paradigms under two different traffic classes UDP and TCP for AAL2R. Figure 6 shows the throughput AAL2R using TCP. The throughput of UDP simulation is shown in figure 7.

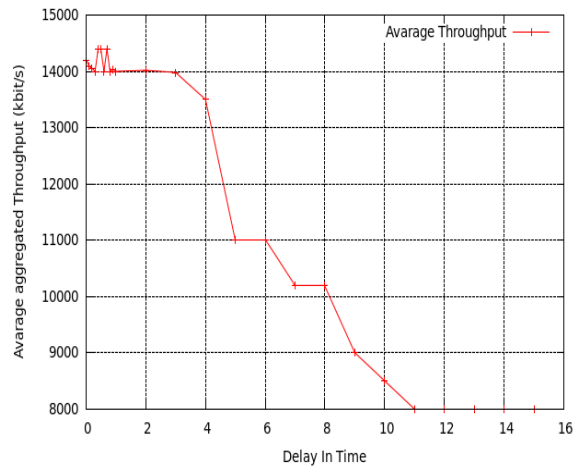

Figure 6. Throughput of AAL2R using TCP

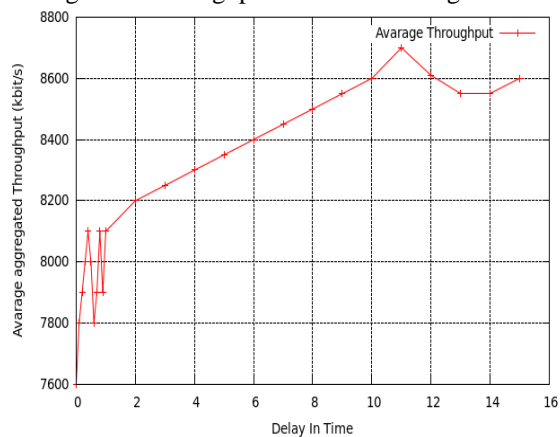

Figure 7. Throughput of AAL2R using UDP

## 5. CONCLUSIONS

In this paper we have studied and elaborated two packet forwarding schemes AAL2R and GSR. AAL2R considers both packet aggregation and packet forwarding but the GSR considers only the packet forwarding strategies. The performance of the resulting network by implementing both algorithms, has been measured by calculating some network influencing parameters like packet delivery ratio, packet loss ratio and throughput. By implementing these algorithms the simulation study shows that that AAL2R has high packet delivery ratio than that of GSR. As the throughput of the network is directly proportional to the packet delivery ratio, hence the AAL2R has also high network throughput over GSR.

It can be concluded that in multi radio multi channel wireless mesh network AAL2R is performing good as it is considering the packet aggregation which directly increases the network capacity by enabling the node to transmit multiple packet in a single time interval. But GSR considers only the packet forwarding which limits the network overhead and capacity. Also AAL2R performs immensely while handling the interference with the presence of heavy traffic as compared to GSR.

## ACKNOWLEDGEMENTS

We would like to thank our parents, teachers and colleagues for their inspiration which helps us to initiate the research work and also for the acceleration during the work. We also like to extend our thanks to everyone from our institute side whose help and expertise advice greatly assist in this research work. We would also like show our immense gratitude to everyone who share their pearls of wisdom with us during the course of work.

**Authors**

**Ms Adyasha Behera**

Ms Adyasha Behera has successfully completed her M.Tech in Information Technology from College of Engineering and Technology, Odisha in 2014. Her research interests include Wireless Mesh Network, Mobile Computing and Programming Language.

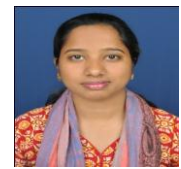

**Mr Amrutanshu Panigrahi**

Mr Amrutanshu Panigrahi has successfully completed his M.Tech in Information Technology from College of Engineering and Technology Odisha in 2014. His research interests include Mobile Computing, Wireless Mesh Network, Distributed System, Real-time operating system and Network security.

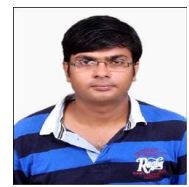